\documentclass[aps,prd,twocolumn,showkeys,superscriptaddress, floatfix]{revtex4-2}

\usepackage{amsmath,amssymb}
\usepackage{graphicx}

\newskip\humongous \humongous=0pt plus 1000pt minus 1000pt

\newif\ifdtup

\relax

\newcommand{\bea}{\begin{eqnarray}}
\newcommand{\eea}{\end{eqnarray}}

\newcommand{\nn}{\nonumber}

\begin{document}
\title{Has Telescope Array Discovered Electroweak Monopole?}
\bigskip
\author{Y. M. Cho}
\email{ymcho0416@gmail.com}
\affiliation{School of Physics and Astronomy, 
Seoul National University, Seoul 08826, Korea}
\affiliation{Center for Quantum Spacetime, Sogang 
University, Seoul 04107, Korea}  
\author{Franklin H. Cho}
\email{cho.franklin@qns.science}
\affiliation{Center for Quantum Nano Science,
Ewha Woman's University, Seoul 03766, Korea}

\begin{abstract}
We propose the ultra high energy cosmic ray recently 
detected by Telescope Array to be the electroweak 
monopole, and present theoretical arguments which support
this. This strongly motivates the necessity for 
the ``cosmic" MoEDAL experiment which could back up 
our proposal. To confirm this we propose Telescope Array 
to measure the magnetic charge of the ultra high energy 
cosmic ray particles with SQUID.  
\end{abstract}

\keywords{ultra high energy cosmic ray, Greisen-Zatsepin-Kuzmin (GZK) limit, electroweak monopole as ultra high energy cosmic 
ray, enhanced Chrenkov radiation of electroweak monopole, cosmological production of electroweak monopole, remnant electroweak monopole density}

\maketitle

Recently the Telescope Array Group (TAG) has announced 
the detection of an ultra high energy cosmic ray (UHECR) 
of energy 244 EeV ($244 \times 10^{18}$ eV) \cite{ta}. 
This is the most recent confirmation of the existence 
of the UHECR particles, following the 320 EeV particle 
in 1991, 213 EeV particle in 1993, and 280 EeV particle 
in 2001 \cite{uhecr}. This tells that the UHECR particles 
exist in nature. 

The TAG (and similar previous) report is very interesting 
and remarkable in two aspects. First, the energy of 
the cosmic ray exeeds the Greisin-Zatsepin-Kuzmin 
(GZK) energy limit \cite{gzk}. Second, the arrival direction 
of the UHECR implies that it came from 
the Local Void. This is puzzling, because there are 
few known particles in nature which could produce 
such UHECR. 

A natural candidate for the UHECR is the proton, 
but it is very difficult for proton to generate 
such high energy. A relativistic proton moving through 
the cosmic microwave background, after collision 
with the 3 K microwave photons becomes $\Delta^*$ and 
decays to nucleons and pions,
\begin{gather}
p + \gamma \rightarrow \Delta^* 
\rightarrow N + \pi.
\label{pgamma}
\end{gather}	
And the mean free path for this process is known to be about 
6 Mpc. This resonant scattering degrades the energy of 
the relativistic protons and prevent them to acquire 
the energy above $5 \times 10^{19}$ eV. This is the GZK 
limit \cite{gzk}. This implies that, if the UHECR of 
TAG is proton, it should have originated nearby, or 
should have energy far above the GZK limit. But these possibilities seems unlikely. 

The other point is that the UHECR observed by TAG 
appears to come from the void, which suggests that 
the origin of this UHECR is not galactic center or other astronomical objects like the neutron stars. This is 
another puzzling feature of this UHECR \cite{ta}. 
 
From this we may conclude that the UHECR of TAG is 
not likely to be an ultra relativistic proton, or any 
known particle produced by the astronomical objects.
This lack of a possible explanation for the UHECR is 
disappointing, and it has been suggested that this could 
be due to ``an incomplete knowledge of particle 
physics" \cite{ta}. {\it The purpose of this Letter is 
to argue that the UHECR observed by TA could be the remnant electroweak monopole produced in the early universe during 
the electroweak phase transition. We present theoretical 
reasons to support this, and discuss how to confirm this 
proposal experimentally.}    

The proposition that the monopoles could be the source 
of the UHECR is not new \cite{nc}. Obviously the monopole, 
if exists, becomes an ideal candidate for the UHECR
particle. It has the strong magnetic interaction, 
stronger than the electric interaction of proton by 
the factor $1/\alpha$. It has the absolute stability guaranteed by the $\pi_2(S^2)$ monopole topology, which 
is required for the UHECR particles. Moreover, it may 
have the mass much heavier than the proton. For this 
reason it has been asserted that the grand unification monopole could generate the UHECR \cite{app}. But in 
this paper we argue that the electroweak (``Cho-Maison") monopole could be the UHECR particles. To do that we 
must understand why the electroweak monopole, not others, 
should be viewed as the UHECR of TAG.

With the advent of the Dirac monopole, the magnetic 
monopole has become an obsession in physics \cite{dirac,cab}. After the Dirac monopole we have had the Wu-Yang monopole \cite{wu}, the 'tHooft-Polyakov monopole \cite{thooft}, 
the grand unification monopole \cite{dokos}, and the electroweak monopole \cite{plb97,yang}. But the electroweak monopole stands out as the most realistic monopole that 
exists in nature and could actually br detected by 
experiment \cite{epjc15,ellis,bb,pta19,epjc20,gv}. 

This is because the Dirac monopole in electrodynamics transforms to the electroweak monopole after 
the unification of the electromagnetic and weak interactions, and the Wu-Yang monopole in QCD becomes unobservable 
after the monopole condensation which confines the color. Moreover, the 'tHooft-Polyakov monopole exists only in 
a hypothetical theory, and the grand unification monopole which could have been amply produced at the grand 
unification scale in the early universe probably has 
become completely irrelevant at present universe after 
inflation. 

Unlike other monopoles the electroweak monopole has 
the following unique features \cite{plb97}. First, 
the magnetic charge is not $2\pi/e$ but $4\pi/e$, twice that of the Dirac monopole. This is because the period 
of the electromagnetic U(1) subgroup of the standard 
model is $4\pi$, since the electromagnetic U(1) comes 
from the U(1) subgroup of SU(2). Second, the mass is 
of the order of several TeV, probably between 4 to 10 TeV. This is because the mass basically comes from the same Higgs mechanism which makes the W boson massive, except that here the coupling is magnetic (i.e., $4\pi/e$). 
This makes the monopole mass $1/\alpha$ times heavier 
than the W boson mass. In spite of this, the size of 
the monopole is set by the weak boson masses. This is because the monopole has the Higgs and W boson dressing which has the exponential damping fixed by the weak boson masses. Third, this is the monopole which exists within (not beyond) the standard model as the electroweak generalization of the Dirac monopole, as a hybrid between Dirac and 'tHooft-Polyakov monopoles. Finally, this monopole is absolutely stable. The topological stability must be obvious, but it also has the dynamical 
stability \cite{gv}.  

The importance of the electroweak monopole comes from
the fact that it must exist if the standard model is 
correct. This means that the discovery of the monopole, 
not the Higgs particle, should be viewed as the final 
(and topological) test of the standard model. Moreover, 
if discovered, it will become the first topologically 
stable magnetically charged elementary particle in 
the history of physics \cite{epjc15}. Furthermore, 
the monopoles produced in the early universe could 
play important roles in physics, in particular in
cosmology \cite{pta19}. Indeed, when coupled to gravity, they automatically turn to the primordial magnetic black holes which could account for the dark matter, become 
the seed of stellar objects and galaxies, creating the large scale structures of the universe. As importantly, they could generate the intergalactic magnetic field,
and could be the source of the ultra high energy cosmic 
rays \cite{pta19}. 

This makes the experimental detection of the electroweak 
monopole a most urgent issue after the discovery of 
the Higgs particle. For this reason MoEDAL and ATLAS 
at CERN are actively searching for the monopole \cite{medal1,medal2,atlas}. Since the electroweak monopole 
has unique characteristics, they could detect the monopole without much difficulty, if LHC could produce them. 
If the monopole mass exceeds 7 TeV, however, the present 
14 Tev LHC may not be able to produce the monopoles. 
In this case we may have to wait for the next upgrading 
of LHC, or else look for the remnant monopoles produced 
in the early universe during the electroweak phase 
transition.

How can we detect the remnant electroweak monopoles?
Obviously we could design a ``cosmic" MoEDAL experiment 
located at high mountains to detect them, and this is 
in planning now. At this point one might wonder if 
the underground experiments like IceCube or Antares 
could be helpful. Unfortunately the underground 
experiments may have difficulty to detect them because 
the penetration length of the monopole in matters is 
expected to be very short (of the order of 10 meters 
in aluminum) because of the strong magnetic 
interaction \cite{icube,anta}. 

Another way to detect the electroweak monopole is 
to detect UHECR particles which could be generated by 
the remnant monopoles, using the cosmic ray experiments. 
The primary purpose of this type of experiments, 
of course, is to detect the high energy cosmic rays
(not the monopoles). Nevertheless this type of 
experiments could be used to detect the remnant 
monopoles, and in this connection the TA experiment 
could play an important role. The question here is why 
and how the remnant electroweak monopoles could be 
identified as the UHECR particles. Now, we are ready 
to discuss how they can become the UHECR particles.

To see this we first notice that the remnant electroweak 
monopoles could easily acquire the energy above the GZK 
limit from the intergalactic magnetic field. Since 
the average intergalactic magnetic field $B$ is 
about $3 \times 10^{-6}$ G with the coherent length $L$
of the order of 300 pc, we could estimate the monopole 
energy gain traveling through the intergalactic magnetic 
field to be \cite{pta19}
\begin{gather}
\Delta E \simeq \frac{4\pi}{e}~B L 
\simeq 1.2 \times 10^{20}~\text{eV}. 
\label{me}
\end{gather} 
Moreover, we can easily show that the monopole energy 
loss due to the linear acceleration is completely 
negligible. This confirms that they could acquire 
the energy above the GZK limit without any problem.

Moreover, unlike the proton, the monopole retains 
its energy traveling through the cosmic microwave  
background. This is because the photon-monopole
scattering cross section is givden by the classical 
Thompson scattering cross section, which is many 
orders of magnitude down the photon-proton cross
section described by (\ref{pgamma}).  

Notice that (\ref{me}) is independent of the monopole 
mass. So, depending on the mass the monopole (even 
with the above energy) could be relativistic or non-relativistic. For example, if it is the grand 
unification monopole of mass of $10^{17}$ GeV, it 
becomes non-relativistic and thus can not generate relativistic secondaries in the cosmic ray shower. 
In contrast, (\ref{me}) makes the electroweak monopole 
with mass of $M_W/\alpha$ extremely relativistic, 
so that it could easily generates the relativistic 
showers. And in reality we do have the relativistic 
showers in the UHECR. This strongly indicates that 
the UHECR could be the electroweak monopole. In 
the following we will assume the monopole mass to be 
$M=M_W/\alpha$ for simplicity.   

How is the electromagnetic shower of the electroweak 
monopole? The magnetic field of the monopole 
$B=(4\pi/e) \hat r/r^2$, when boosted to the energy (\ref{me}), 
generates the electric field 
$\vec E = \gamma \beta \vec v \times \vec B$. So, 
the electromagnetic energy loss of the relativistic electroweak monopole should be similar to that of 
a heavy charged particle of mass $M_W/\alpha$ with 
similar $\gamma$ factor (for our monopole with energy (\ref{me}), we have $\gamma \geq 10^{7}$) and charge $1/\alpha$. Moreover, 
the Cherenkov radiation of 
the monopole is enhanced by the factor $(1/\alpha)^2$, compared to that of the proton. This enhanced 
Cherenkov radiation is an important feature of the UHECR 
generated by the electroweak monopole, which could be 
useful in identifying the UHECR particle as the electroweak monopole. 

As for the hadronic shower of the electroweak monopole, 
notice that the maximum fraction of the energy 
transferred from our monopole of mass $M$ to the target particles with mass $m$ is given by \cite{app}
\begin{gather}
\frac{E'_m}{E_M} =\frac{m}{E_M}
\Big(\frac{4E_M^2-2M^2+m^2}{4mE_M +2 M^2} \Big).
\end{gather}  
Now, for our case we have $E_M^2 >> M^2 >> m^2$, so that
\begin{gather}
\frac{E'_m}{E_M} \simeq \frac{2mE_M}{2mE_M +M^2} 
\simeq 1.
\label{elm}
\end{gather} 
This should be compared to the maximum energy transfer 
of the proton (with $M \simeq m$)
\begin{gather}
\frac{E'_m}{E_M} = 1 -\frac{m}{2E_M} \simeq 1,
\label{elp}
\end{gather} 
which is not so different from (\ref{elm}). 
From this we may conclude that our monopoles transfer 
most of the energy in the first forward hadronic scattering, 
and thus produce an air shower resembling a typical 
hadron initiated shower. This implies that trying to 
identify the UHECR with hadronic shower pattern may 
not be a wise strategy.   

Now, we have to discuss the remnant electroweak 
monopole density at present universe. This is 
a complicated issue, but fortunately this has already 
been studied before \cite{pta19}. To summarize 
the results we start from the temperature dependent 
effective action of the standard model
\begin{gather}
V_{eff}(\rho) =V_0(\rho) -\frac{C_1}{12\pi} \rho^3 T
+\frac{C_2}{2} \rho^2 T^2 
-\frac{\pi^2}{90} g_* T^4  
+\delta V_T,  \nn\\
V_0(\rho)=\frac{\lambda}{8}(\rho^2-\rho_0^2)^2 ,  \nn\\
C_1=\frac{6 M_W^3 + 3 M_Z^3}{\rho_0^3}\simeq 0.36,   \nn\\
C_2=\frac{4M_W^2 +2 M_Z^2 +M_H^2+4m_t^2}{8\rho_0^2} 
\simeq 0.37,   
\label{epot}
\end{gather}
where $V_0$ is the zero-temperature potential, $g_*$ is 
the total number of distinct helicity states of 
the particles with mass smaller than $T$, $C_1$ and 
$C_2$ are the contributions from the weak bosons and 
fermions, $m_t$ is the top quark mass, and $\delta V_T$ 
is the slow-varying logarithmic corrections and the lighter 
quark contributions which we will neglect from now on.

\begin{figure}
\includegraphics[height=4.5cm, width=7cm]{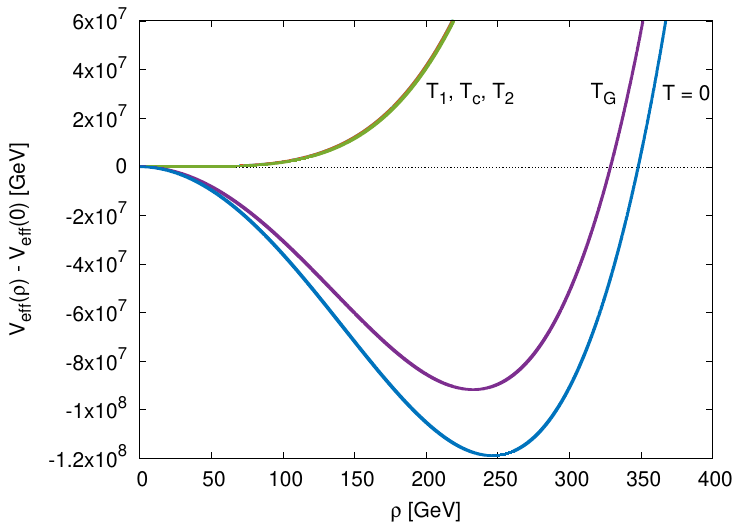}
\caption{\label{tpot} The temperature dependent effective potential (\ref{epot}) at various temperatures, where 
$T_G$ is the Ginzburg temperature. Notice that the potential at $T_1,~T_c,~T_2$ are almost indistinguishable.}
\end{figure}

The potential has three local extrema at 
\begin{gather}
\rho_s=0,   \nn\\
\rho_{\pm}(T)=\Big\{\frac{C_1}{4\pi \lambda}
\pm \sqrt{\Big(\frac{C_1}{4\pi \lambda} \Big)^2
+\frac{\rho_0^2}{T^2} -\frac{2C_2}{\lambda}} \Big\}~T^.
\label{rext}
\end{gather}
The first extremum $\rho_s=0$ represents the Higgs vacuum 
of the symmetric (unbroken) phase, the second one $\rho_-(T)$ represents the local maximum, and the third one $\rho_+(T)$ represent the local minimum Higgs vacuum of the broken phase. It is charactrized by three temperatures, 
\begin{gather}
T_1 =\frac{4 \pi \lambda}{\sqrt{32\pi^2\lambda C_2-C_1^2}}
~\rho_0 \simeq 146.13~\text{GeV},  \nn\\
T_c= \sqrt{\frac{18}{36\pi^2 \lambda C_2- C_1^2}} 
~\pi \lambda \rho_0  \simeq 146.09 ~{\rm GeV},   \nn\\
	T_2=\sqrt{\frac{\lambda}{2C_2}}~\rho_0 
\simeq 145.82~\text{GeV}.  
\end{gather}
Above $T_1$ only $\rho_s=0$ becomes the true vacuum of 
the effective potential, and the electroweak symmetry 
remains unbroken. At the critical temperature $T_c$, 
the two vacua $\rho_s$ and $\rho_+$ are degenerate 
and the electroweak phase transition starts. At $T_2$ 
we have only one vacuum $\rho_+$ with $\rho_0=\rho_-$, 
and the phase transition ends. So, in principle 
the electroweak phase transition is the first order. 
Since $T_1$, $T_c$, and $T_2$ are very close, however, 
the phase transition practically becomes the second 
order \cite{pta19}. The effective potential (\ref{epot}) 
is graphically shown in Fig. \ref{tpot}.

The monopole production in the second order phase transition is supposed to be described by the Kibble-Zurek mechanism, 
so that the monopole production start from $T_c$. However, 
the thermal fluctuations of the Higgs vacuum which create 
the seed of the monopoles continue till the universe cools
down to the Ginzburg temperature $T_G \simeq 57.6~\text{GeV}$, where the monopole production stops\cite{pta19,gin}. 
The Ginzburg temperature is shown in Fig. \ref{tpot}.  
So, the electroweak monopole formation takes place 
between $T_c$ and $T_G$, or in average around $T_i$,
\begin{gather}
T_i= \frac{T_c+T_G}{2} \simeq  101.7~\text{GeV}.
\label{itemp}
\end{gather}
In time scale, we can say that the electroweak monopole 
production start from $1.8 \times 10^{-11} sec$ to 
$1.2 \times 10^{-10} sec$ after the big bang for 
the period of $10.3 \times 10^{-11}~sec$, or around 
$3.5 \times 10^{-11} sec$ after the big bang in average.

Two important parameters of the electroweak phase transition
are the temperature dependent Higgs boson mass $\bar M_H$ which determine the correlation length $\xi=1/ \bar M_H$ and 
the W-boson mass $\bar M_W$ which determines the monopole 
mass $\bar M \simeq \bar M_W /\alpha$. The Higgs boson acquires its minimum mass 5.54 GeV at $T=T_c$ which approaches to 
the zero temperature value 125.2 GeV as the universe cools 
down. The W-boson which is massless before the symmetry 
breaking becomes massive toward the value 6.76 GeV at $T_c$ 
and 73.2 GeV at $T_G$. This tells that the infant monopole 
masses at $T_c$ and $T_G$ are around 1.4 TeV and 10 TeV 
(with the adolescent mass 10.7 TeV). 

According to the Kibble-Zurek mechanism the initial monopole density is determined by the mean value of two correlation 
lengths at $T_c$ and $T_G$ \cite{kibble,zurek},
\begin{gather}
\xi_i=\frac{\xi(T_c)+\xi(T_G)}{2}\simeq 9.4 
\times 10^{-16}~\text{cm}.
\label{incol}
\end{gather} 
From this we can estimate the initial density of 
the monopoles to be $n_i \simeq T_i^3/\xi_i^3 
\simeq  0.2~T_i^3$. This estimate, however, has a defect 
that the energy within one correlation volume is not 
enough to provide the monopole mass. This is because 
the monopole mass is $M_W/\alpha$, but the size is fixed 
by $M_W$. A natural way to cure this defect is to adopt 
a new correlation length $\bar \xi_i$ which satisfies 
the energy constraint,
\begin{gather}
\bar \xi_i = \Big(\frac{1}{\alpha} \Big)^{1/3} \xi_i
\simeq 5.16 \times \xi_i.
\label{ec1}
\end{gather}
With this the initial monopole density is given by
\begin{gather}
n_i \simeq \frac{T_i^3}{\bar \xi_i^3}
=\alpha \times \frac{T_i^3}{\xi_i^3}
\simeq  1.5 \times 10^{-3}~T_i^3.
\label{imd}
\end{gather}
This is smaller than the Kibble-Zurek estimate by 
the factor $\alpha$.  

\begin{figure}
\includegraphics[height=4.5cm, width=7cm]{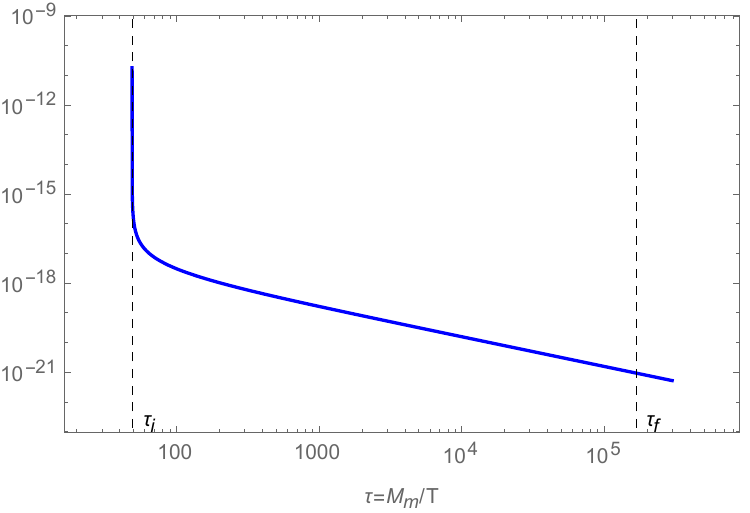}
\caption{\label{mden} The cosmic evolution of 
the electroweak monopole density $n_m/T^3$ against 
$\tau=M_m/T$.}
\end{figure}

To determine the remnant monopole density at present 
universe, however, we have to know how it evolves in 
cosmology. The evolution of the monopoles is determined 
by the Boltzmann's equation \cite{pres}
\begin{gather}
\frac{d n}{dt} + 3 H n = -\sigma n^2,
\label{Beq}
\end{gather}
where $H$ is the Hubble parameter and 
$\sigma \simeq 1/3\alpha T^2$ is the monopole annihilation 
cross section. The solution of the evolution equation is 
shown in Fig. \ref{mden}. Notice that the monopoles, as soon 
as produced, quickly annihilate each other. This is because 
at the initial stage of the monopole production, the capture radius of the monopole and anti-monopole is much bigger than 
the correlation length. This quickly reduces the initial 
monopole density by the factor $10^{-6}$. Moreover, 
the final value of the monopole density becomes independent 
of the initial value and approaches to
$n \rightarrow 18.25~(T/M_P) \times T^3$ regardless of 
the initial condition, where $M_P$ is the Planck 
mass \cite{pres}. And the monopole-antimonopole annihilation 
ceases at the temperature $T_f \simeq 60~{\text {MeV}}$. 
This is below the muon decoupling temperature, which tells 
that the annihilation of the monopoles continues very long 
time. 

Below $T_f$ the monopoles are free streaming, and we can 
estimate the remnant monopole density at the present 
universe. The monopole density at $T_f$ becomes 
\begin{gather}
n_f \simeq 0.9 \times 10^{-19}~T_f^3,
\label{fmden}
\end{gather}
which is much lower than the initial density given by 
(\ref{imd}). The number of monopole within the co-moving 
volume is conserved thereafter. But they still interact 
with the electron pairs in the hot plasma before decouple 
around $T_d\simeq 0.5~\text{MeV}$, when the electron pairs 
disappear and the interaction rate becomes less than 
the Hubble expansion rate. Since the decoupling temperature 
of the electroweak monopole is much less than the monopole 
mass, the free streaming monopoles just after the decoupling 
start as completely non-relativistic. But eventually they 
become extremely relativistic by the acceleration of 
the intergalactic magnetic field.  
 
Assuming that the expansion is adiabatic, the current 
number density and the energy density of the monopole 
is given by \cite{pta19}
\begin{gather}
n_0= \frac{g_0}{g_f}~\Big(\frac{T_0}{T_f} \Big)^3~n_f, \nn\\
\rho_m = n_0~M 
= \frac{g_0}{g_f}~\Big(\frac{T_0} {T_f} \Big)^3~n_f~M,
\label{mden0}
\end{gather}
where $T_0 = 2.73~\text{K}=2.35\times 10^{-13}~\text{GeV}$ 
is the temperature of the universe today and $g$ is 
the effective number of degrees of freedom in entropy. 
So we have the current density of monopole 
\begin{gather}
\Omega_m~h^2 =\frac{\rho_m}{\rho_c}~h^2
\simeq 1.2 \times 10^{-10},
\label{denpara}
\end{gather}
where $\rho_c$ is the critical density of present universe 
and $h\simeq 0.678$ is the scaled Hubble constant in 
the unit $H_0/(100~\text{km}~\text{s}^{-1}~\text{Mpc}^{-1})$.
This is about $1.3 \times 10^{-9}$ of the baryon density.
This assures that the electroweak monopole does not alter 
the standard big bang cosmology in any significaly way.

In terms of the number density, this translates to 
about $6.1 \times 10^{-5} /~\text{Km}^3$, or about 
$2.3 \times 10^{-13}~n_b$, where $n_b$ is the number 
density of the baryons. This is roughly $10^5$ times 
bigger than the monopole density set by the Parker 
bound \cite{parker}, which implies that (\ref{mden0})  
is an over estimation. 

Actually there are reasons that the real remnant monopole 
density could be much less \cite{pta19}. First, as 
the only heavy stable particle with mass about $10^4$ 
times heavier than the proton, they can easily 
generate the density perturbation and might have been 
buried in galactic centers. Second, they have a very 
short penetration length in matters, so that most of 
them could have been trapped and filtered out by 
the stellar objects after collision with them. Third 
(and most importantly), when coupled to gravity, they 
automatically turn to Reissner-Nordstrom type black 
holes and become the premordial magnetic black holes. 
This strongly implies that indeed (\ref{mden0}) could 
be made consistent with the Parker bound.

With this remark we can confidently say that the UHRCR 
particle observed by TA could be the remnant electroweak 
monopole produced in the early universe, as one of us 
has pointed out in an earlier work \cite{pta19}. In 
particular, our estimate of the monopole density appears 
to be consistent with the UHECR event rate at TA, and 
it comes from the void as TA indicated. 

Unfortunately, this proposition could only be confirmed 
indirectly at TA at the moment, with the enhanced Cherenkov 
radiation of the monopole. To confirm this directly, 
TA should be able to measure the magnetic charge of 
the UHECR. In principle this could be done by installing 
SQUID to each of the surface detectors at TA. We hope that 
TAG could measure the magnetic charge of the UHECR with 
SQUID in the near future. The details of our discussions 
will be published in a separate paper \cite{cho}.      
  
{\bf ACKNOWLEDGEMENT}

The work is supported in part by the National Research 
Foundation funded by the Ministry of Education
(Grant 2018-R1D1A1B0-7045163), the Ministry of Science 
and Technology (Grant 2022-R1A2C1006999), and by Center 
for Quantum Spacetime, Sogang University, Korea.

\end{document}